\documentclass[aps,prl,reprint,amsfonts,superscriptaddress]{revtex4-1}

\usepackage{graphicx}

\def\Ef{$E_{\rm F}$}
\def\Eb{$E_{\rm B}$}

\def\Gbar{$\overline{\Gamma}$}

\def\GbarXbar{$\overline{\Gamma}$-$\overline{\rm X}$}

\def\BiTe{Sb$_2$Te$_3$}
\def\BiTe{Bi$_2$Te$_3$}
\def\BiSe{Bi$_2$Se$_3$}
\def\BiX{Bi$_2$X$_3$}
\def\aSn{$\alpha$-Sn}

\def\SbTe{Sb$_2$Te$_3$}

\def \Z{$\mathbb{Z}_2$}

\begin{document}


\title{Elemental Topological Insulator with a Tunable Fermi Level:\\Strained $\alpha$-Sn on InSb(001)}

\author{A. Barfuss}
\affiliation{\mbox{Physikalisches Institut und R\"ontgen Center for Complex Materials Systems, Universit\"at W\"urzburg, 97074 W\"urzburg, Germany}}
\author{L. Dudy}
\affiliation{\mbox{Physikalisches Institut und R\"ontgen Center for Complex Materials Systems, Universit\"at W\"urzburg, 97074 W\"urzburg, Germany}}
\author{M. R. Scholz}
\affiliation{\mbox{Physikalisches Institut und R\"ontgen Center for Complex Materials Systems, Universit\"at W\"urzburg, 97074 W\"urzburg, Germany}}
\author{H. Roth}
\affiliation{\mbox{Physikalisches Institut und R\"ontgen Center for Complex Materials Systems, Universit\"at W\"urzburg, 97074 W\"urzburg, Germany}}
\author{P. H\"opfner}
\affiliation{\mbox{Physikalisches Institut und R\"ontgen Center for Complex Materials Systems, Universit\"at W\"urzburg, 97074 W\"urzburg, Germany}}
\author{C. Blumenstein}
\affiliation{\mbox{Physikalisches Institut und R\"ontgen Center for Complex Materials Systems, Universit\"at W\"urzburg, 97074 W\"urzburg, Germany}}
\author{\mbox{G. Landolt}}
\affiliation{Swiss Light Source, Paul-Scherrer-Institut, 5232 Villigen, Switzerland}
\affiliation{Physik-Institut, Universit\"at Z\"urich-Irchel, 8057 Z\"urich, Switzerland}
\author{J. H. Dil}
\affiliation{Swiss Light Source, Paul-Scherrer-Institut, 5232 Villigen, Switzerland}
\affiliation{Physik-Institut, Universit\"at Z\"urich-Irchel, 8057 Z\"urich, Switzerland}
\author{N. C. Plumb}
\affiliation{Swiss Light Source, Paul-Scherrer-Institut, 5232 Villigen, Switzerland}
\author{M. Radovic}
\affiliation{Swiss Light Source, Paul-Scherrer-Institut, 5232 Villigen, Switzerland}
\author{A. Bostwick}
\affiliation{Advanced Light Source, Lawrence Berkeley National Laboratory, Berkeley, CA 94720, USA}
\author{E. Rotenberg}
\affiliation{Advanced Light Source, Lawrence Berkeley National Laboratory, Berkeley, CA 94720, USA}
\author{A. Fleszar}
\affiliation{\mbox{Institut f\"ur Theoretische Physik und Astronomie, Universit\"at W\"urzburg, 97074 W\"urzburg, Germany}}
\author{\mbox{G. Bihlmayer}}
\affiliation{\mbox{Peter Gr\"unberg Institute and Institute for Advanced Simulation, Forschungszentrum J\"ulich, 52425 J\"ulich, Germany}}
\author{D. Wortmann}
\affiliation{\mbox{Peter Gr\"unberg Institute and Institute for Advanced Simulation, Forschungszentrum J\"ulich, 52425 J\"ulich, Germany}}
\author{G. Li}
\affiliation{\mbox{Institut f\"ur Theoretische Physik und Astronomie, Universit\"at W\"urzburg, 97074 W\"urzburg, Germany}}
\author{W. Hanke}
\affiliation{\mbox{Institut f\"ur Theoretische Physik und Astronomie, Universit\"at W\"urzburg, 97074 W\"urzburg, Germany}}
\author{R. Claessen}
\affiliation{\mbox{Physikalisches Institut und R\"ontgen Center for Complex Materials Systems, Universit\"at W\"urzburg, 97074 W\"urzburg, Germany}}
\author{J. Sch\"afer}
\affiliation{\mbox{Physikalisches Institut und R\"ontgen Center for Complex Materials Systems, Universit\"at W\"urzburg, 97074 W\"urzburg, Germany}}
\email[]{Joerg.Schaefer@physik.uni-wuerzburg.de}

\date{\today}

\begin{abstract}
We report on the epitaxial fabrication and electronic properties of a topological phase in strained \aSn\ on InSb. The topological surface state forms in the presence of an unusual band order not based on direct spin-orbit coupling, as shown in density functional and GW slab-layer calculations. Angle-resolved photoemission including spin detection probes experimentally how the topological spin-polarized state emerges from the second bulk valence band. Moreover, we demonstrate the precise control of the Fermi level by dopants.
\end{abstract}


\maketitle


Two- and three-dimensional (2D, 3D) topological insulators (TIs) owe their conductance to spin-polarized edge or surface states, respectively \cite{Qi:2011hb,Kane:2005hl,Koenig:2007hs,Fu:2007wc}. Assuming time-reversal symmetry (TRS), an inversion between occupied and unoccupied bands of different parity is usually caused by strong spin-orbit (SO) coupling, which alters the topological invariant \Z\ from $\nu$=0 (trivial insulator) to $\nu$=1 (topological insulator) \cite{Kane:2005gb,Fu:2007ei}. This gives rise to an odd number of gapless boundary states \cite{Kane:2008cc}. For such a state, TRS demands that electrons moving in opposite directions carry opposite spins, which are therefore protected against backscattering. They promise lossless transport and new avenues for spin-based data processing.

A number of materials were proposed as TIs based on parity inversions \cite{Fu:2007ei}. Some have been investigated by angle-resolved photoemission (ARPES), which can directly probe the topological surface state (TSS) with its two linearly dispersing bands that cross at the Dirac point. Examples are Bi$_{0.9}$Sb$_{0.1}$ \cite{Hsieh:2009if}, \BiSe\ \cite{Xia:2009ii}, \BiTe\ \cite{Chen:2009do}, and \SbTe\ \cite{Pauly:2012gr} (\BiX\ family). A 2D TI is predicted \cite{Bernevig:2006ij} for HgTe quantum wells embedded in CdTe. Closely related to HgTe in terms of band structure is $\alpha$-Sn, which in strained form (together with HgTe) has been proposed as a 3D TI \cite{Fu:2007ei}, however, has not yet been explored. The band order in both high-Z systems (with zero gap unless strained) is rather uncommon: The crystal symmetries together with relativistic corrections give rise to the necessary inversion of band parity - unlike in the \BiX\ family where the SO partner bands are connected by the TSS \cite{Zhang:2009ks}. 

In HgTe/CdTe quantum wells the conductivity of $\sigma=2e^2/h$ is explained by spin-polarized edge channels in the quantum spin Hall effect \cite{Koenig:2007hs,Brune:2010ed,Brune:2012fi}, giving proof of a 2D TI. A crossover to 3D has been reported for thick HgTe films, judged from odd Hall plateaus \cite{Brune:2011hi} as well as ARPES data of a linear surface state \cite{Brune:2011hi,Crauste:2013vx}. However, a direct probe of the spin character by spin-resolved photoemission (SARPES) is missing \cite{Crauste:2013vx}. Moreover, the delicate growth together with the toxicity of Hg makes handling of this system difficult.

The material in focus of the present work, $\alpha$-Sn on InSb, is likewise a heteroepitaxial strained system. The use of InSb as a template for high quality crystals is well established \cite{Farrow:1981il,Mason:1992fe,Magnano:2002uu}. It induces a slight compressive strain of 0.14\% in the diamond lattice of $\alpha$-Sn. The bulk band structure of $\alpha$-Sn/InSb was the subject of a few photoemission reports for the (001) \cite{Hochst:1983hs,Fantini:2000uu} and (111) surface \cite{HernandezCalderon:1985we,Fantini:2000df}. These early studies at low resolution, however, did not address the question of a TSS. To date, experimental and computational demonstrations are still missing.  
 
In this Letter, we report on a 3D topological phase in strained \aSn. By comparing ARPES including spin detection with density functional theory (DFT) and GW quasi-particle calculations, we show that the topological situation emerges from an unusual band order not simply explained by SO coupling. The spin character of the TSS is probed directly, showing a counter-clockwise in-plane rotation for the occupied states of the Dirac cone. In addition, we demonstrate control of the Fermi level by adding dopants. This is the first realization of an elemental TI, which promises ease of fabrication and tuning.

\aSn\ samples have been synthesized \textit{in situ} by molecular-beam epitaxy in ultra-high vacuum on a clean InSb(001) substrate with an additional Te dopant flux, leading to high surface quality as monitored by low-energy electron diffraction. ARPES measurements have been performed with He-I and Ne-I excitation (21.22\,eV and 16.86\,eV), and by tunable synchrotron radiation. Experimental details as well as the numerical procedures are described in the Supplementary Material \footnote{supplementary material}.



\begin{figure}
 	\includegraphics{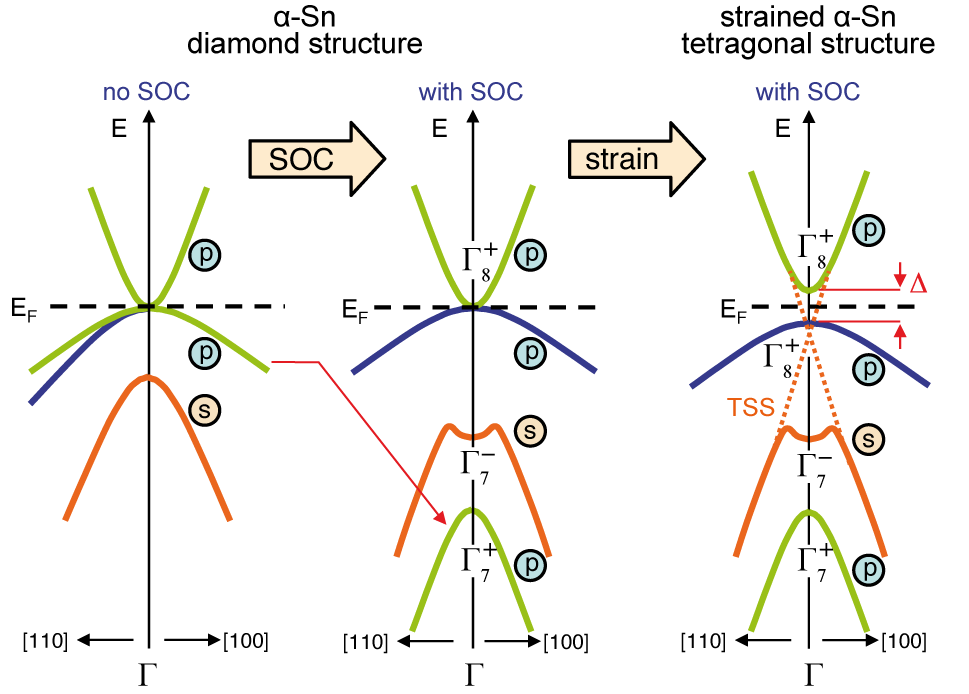}
 	\caption{\label{} Band schematics of \aSn\ for different situations without and with SO coupling (SOC), and with additional strain (double group representations refer to the diamond structure only, but are also used as band labels). Parity inversion exists independent of the presence of SOC as an inherent band feature. The TSS is hosted between second valence band ($\Gamma_7^-$) and conduction band, while the deep SO split-off band ($\Gamma_7^+$) is not involved. Strain opens a small band gap at the $\Gamma$-point.}
  \end{figure}

In \aSn, the band situation is different from the \BiX\ TIs, where the SO coupling drives the parity inversion \cite{Zhang:2009ks}. The unstrained \aSn crystal is a zero-gap material in which the light-hole valence band ($\Gamma_8^+$) and the $\Gamma_7^-$ band (usually the conduction band) are inverted \cite{Groves:1963fy,Pollak:1970jc} compared to other semiconductors in diamond and zinc-blende structure \cite{Harrison:1980}, see band schematic in Fig. 1. The closed gap results from degenerate states at the $\Gamma$-point, where the heavy-hole valence band and the conduction band touch each other at $\Gamma_8^+$. The effect of SO coupling is to impose a strong downward shift (middle panel of Fig. 1), which leads to a deep-lying split-off band ($\Gamma_7^+$). 

This gives rise to the unusual band situation in \aSn: Since the SO split-off band ($\Gamma_7^+$) is even lower than the $\Gamma_7^-$ band, this latter (second) valence band can host the TSS in conjunction with the conduction band of opposite parity. In addition, there is the first valence band ($\Gamma_8^+$) in-between, representing a further modification compared to the \BiX\ family with a direct (SO-driven) topological band pairing. Lattice strain then can lift the band degeneracy at the $\Gamma$-point, as shown in the right panel of Fig. 1 for in-plane compressive strain (i.e. a tetragonal lattice). A small energy gap opens at \Ef\, while the band order remains otherwise unchanged.

The bands for bulk \aSn\ with the appropriate strain (0.14\%) have been derived from a GW quasi-particle calculation (including SO coupling) and are displayed in Fig. 2a). The strain induces a small $\Gamma$-point gap of $\sim$30 meV at the Fermi level. The second valence band - as the candidate to host the TSS - exhibits a weak M-shape and a maximum roughly located at 0.5 eV below \Ef. Interestingly, and unlike in \BiX, this carries a strong s-orbital character (rather than p-character). The third valence band (from SO coupling) is located slightly below.

 \begin{figure}[b]
 	\includegraphics{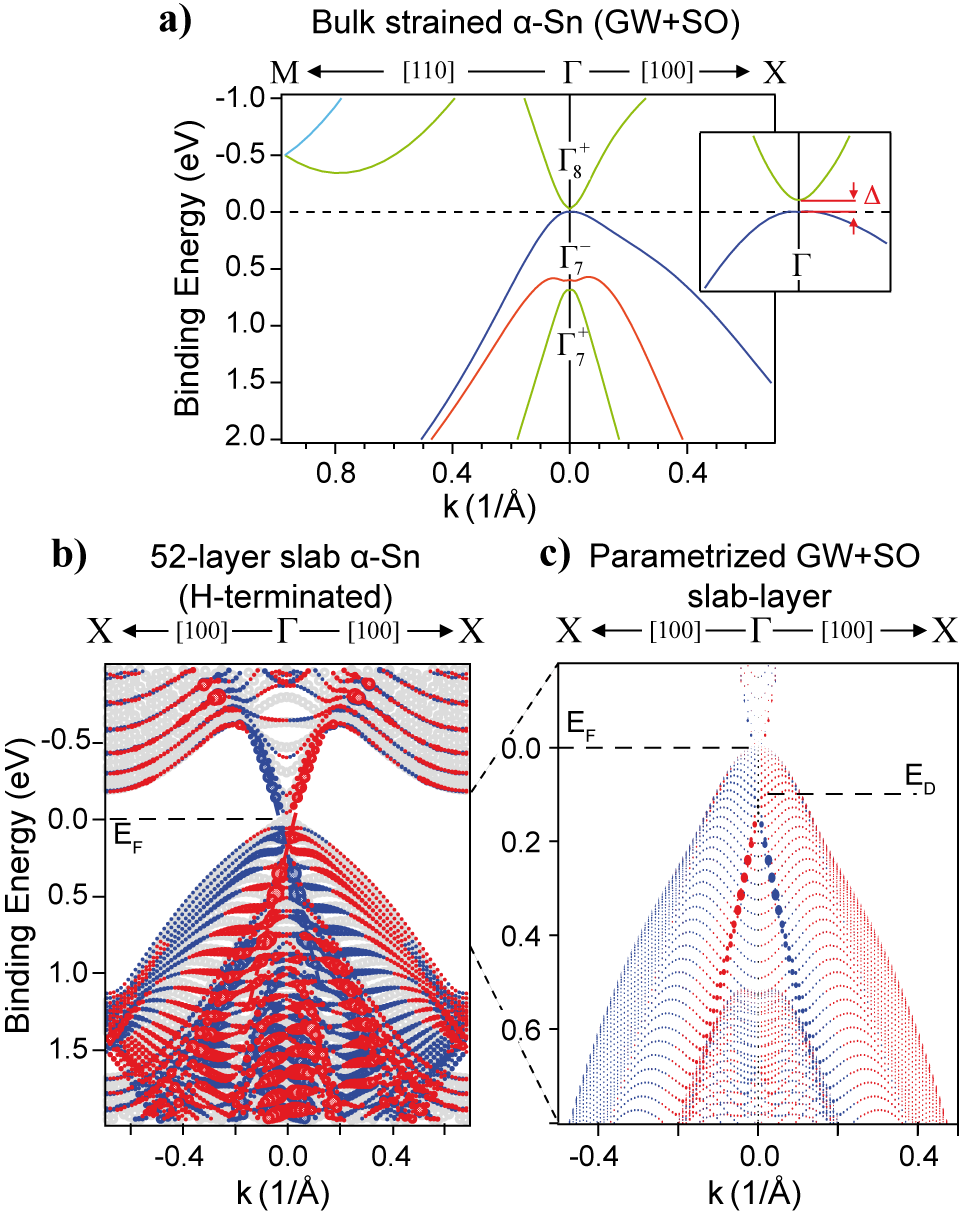}
 	\caption{\label{} a) GW quasi-particle calculation for bulk \aSn\ with realistic strain for InSb substrate (0.14\% compressive). A small gap ($\sim$30 meV) opens at the $\Gamma$-point. b) Slab-layer LDA+U calculation for the (001) surface (52 layers, H-terminated) that shows the TSS with linear dispersion. Opposite spin character is coded in red and blue, respectively, reflecting a counter-clockwise spin orientation around the Dirac cone. c) Close-up view of the TSS bands, based on a slab calculation performed in 256 layers using a tight-binding parametrization of a bulk GW quasi-particle calculation (see Supp. Mat. [25]). The TSS is clearly seen to emerge from the second valence band (colors: spin, symbol size: surface character within topmost 30 layers), with a Dirac point $\sim$ 0.1 eV below \Ef.}
  \end{figure}

In order to generate the TSS, which requires existence of surfaces, we have performed slab-layer calculations, see Fig. 2 b), where two linearly dispersing states (on a bulk band background) are clearly discernible (unlike simple LDA, the LDA+U used here produces the correct band order, as does GW). In addition, we have evaluated the spin character, which shows a clear separation into opposite signs (fully in-plane for the main symmetry lines). The spin helicity is counter-clockwise below the Dirac point, and reversed above. Quantization effects present in small slab models can be avoided using a semi-infinite bulk connection to the surface layers, as in supplementary Fig.\,S1 [25]. A detailed view of the TSS is obtained from GW calculations where a tight-binding parametrization was implemented in a particularly thick slab with 256 layers. Here it becomes apparent that the TSS emerges straight from the second valence band, as assumed above based on the band parity sequence. It gives rise to a Dirac crossing at about 0.1 eV below the Fermi level.


In Fig.\,3a) we present ARPES data of pristine \aSn\  taken with an excitation energy of h$\nu$ = 119\,eV which corresponds approximately to the $\Gamma$-point of the bulk BZ and, thus, compares directly to the band structure calculations of Fig.\,2. We identify the $\Gamma_8^+$ and $\Gamma_7^-$ bands, which appear rather intense at this photon energy. Interestingly, the $\Gamma_7^-$ band looses much of its intensity close to normal emission, i.e., k = 0, which is caused by parity-related matrix elements and, thus, directly reflects the uneven parity at $\Gamma$. However, the linearly dispersing feature seen in our DFT calculations is not resolved at 119\,eV in this sample.

\begin{figure}
 	\includegraphics{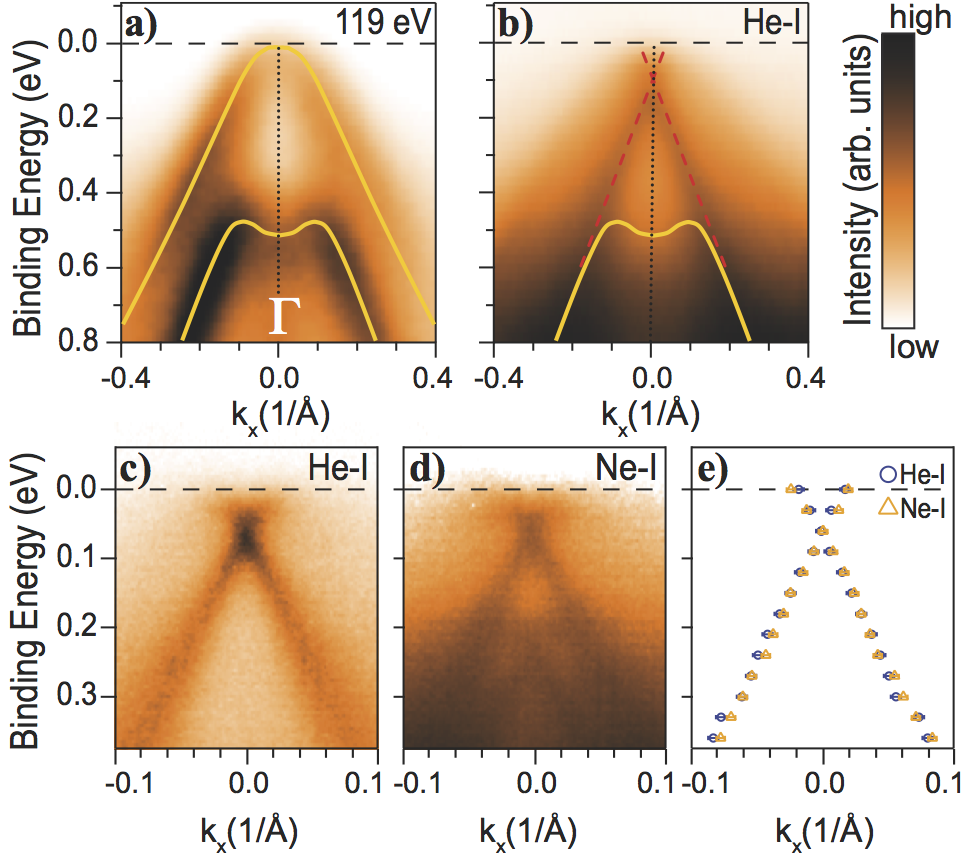}
 	\caption{\label{}a) ARPES measurements on \aSn, performed at 119\,eV, i.e., at the $\Gamma$-point of the BZ resolve the $\Gamma_8^+$ and the $\Gamma_7^-$ valence bands, marked by yellow lines. b) At h$\nu$ = 21.22\,eV (He-I) a linearly dispersing feature is observed and marked by red dashed lines.  (c - e) Photon-energy dependent measurements, i.e., at h$\nu$ = 21.22\,eV (c) and 16.86\,eV (Ne-I) (d) clarify the surface character of the state. Blue circles and orange triangles in e) represent the fitted peak positions from c) and d), respectively. Note that the sample of b) - e) is doped with Te, as described in the text. All measurements taken along the \GbarXbar-direction at room temperature in our home lab (b - e) and at the ALS (a). The faint replicas at higher binding energies in d) are due to a satellite line in the Ne discharge.} 
  \end{figure}
The situation changes drastically at h$\nu$ = 21.22\,eV which gives a better observation condition for the sought-after linear feature (Fig.\,3b). Additionally, the sample of Fig.\,3b) has an increased surface quality as compared to the one of Fig.\,3a), since Te which acts as a surfactant was added during the growth, as discussed below. The change of h$\nu$ adjusts the perpendicular momentum of the photoelectrons k$_z$ close to the X-point of the bulk BZ. The $\Gamma_8^+$-valence band is smeared out to a weak background intensity but, centered around \Gbar\ of the surface BZ, we now find a prominent feature that seems to cross \Ef\ with a rather linear dispersion. In addition, the close-up in Fig.\,3c) suggests a crossing of the two visible branches at a binding energy of $\sim$60\,meV. In agreement with our calculations presented in Fig.\,2, the state emerges out of the $\Gamma_7^-$-valence band, suggesting that it is indeed a TSS. 
\begin{figure}
 	\includegraphics{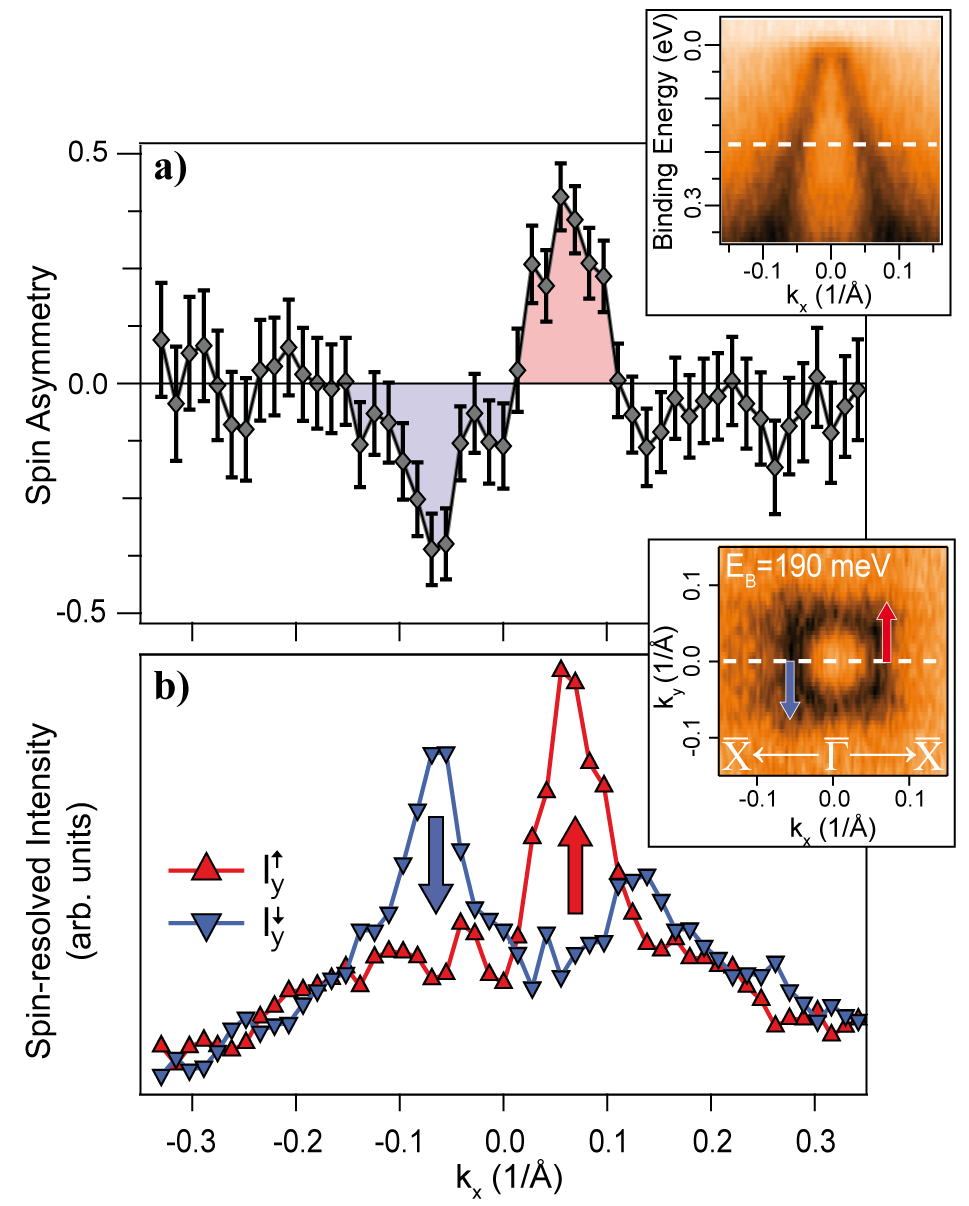}
 	\caption{\label{}SARPES on \aSn. a) Mott polarimetry of an MDC at E$_B$=190\,meV along the \GbarXbar-direction (marked by a white dashed line in both insets), reveals a strong spin-asymmetry that we assign to the surface state. b) The derived spin-resolved intensity distribution gives one highly spin-polarized peak per spin direction, i.e., a \textit{spin down}-peak ($\downarrow$) at  -k$_x$ and a \textit{spin up}-peak ($\uparrow$) at  +k$_x$ (blue and red triangles/lines, respectively). The polarization vector points fully along the y-direction, which gives a counter-clockwise spin helicity as shown in the constant energy surface in the lower inset. All measurements at h$\nu$=19.5\,eV and at T$\sim$20\,K. Spin-resolved and spin-integrated measurements were performed at COPHEE and HRPES endstations at SLS, respectively.}
  \end{figure}
We can experimentally justify this suggestion further, as we characterize the feature to be a 2D surface state by changing the photon energy to h$\nu$ = 16.86\,eV (Fig.\,3d). A surface state should be unaffected by this change as it shows no dispersion perpendicular to the surface. Indeed, as we quantify by an overlay of the fitted momentum distribution curve (MDC) peak positions (Fig.\,3e), the state is essentially unaltered by this change. 

Hence, we find the linear state to satisfy key characteristics of a TSS that are surface localization and metallicity. 
Final evidence is obtained from the spin-polarization of the surface state. For a TI the two branches of the state should show opposite spin-polarization, in agreement with time-reversal symmetry. 
In Fig.\,4, we present our results from spin-resolved photoemission measurements on \aSn. A MDC at \Eb\ = 190\,meV, as marked in the upper inset of Fig.\,4, reveals a strong spin-asymmetry when Mott-polarimetry is applied to the photoelectrons (Fig.\,4a). We observe two features of opposite sign which match very well with the peak positions of the TSS in k-space. This leads to a spin-resolved intensity distribution where the state at k$_x>0$ has spin along +k$_y$ while the state at k$_x<0$ has spin along -k$_y$ (Fig.\,4 b).  
 
While we have measured all three spatial components of the spin vector \cite{Dil:2009ix}, we limit our presentation to the y-component, as the polarization lies, within the error of the measurement, fully in this direction. 
Since the MDC is taken along the k$_x$ direction this means that the spin is oriented perpendicular to the momentum along the high-symmetry direction \GbarXbar. As sketched on the constant energy surface at \Eb\ = 190\,meV (lower inset of Fig. 4) the sign of the spin-vector is such that the helicity is counter-clockwise below the Dirac point, in agreement with our own DFT calculations and findings in other TIs \cite{Souma:2011ci, Pan:2011be, Pauly:2012gr}.

\begin{figure}
 	\includegraphics{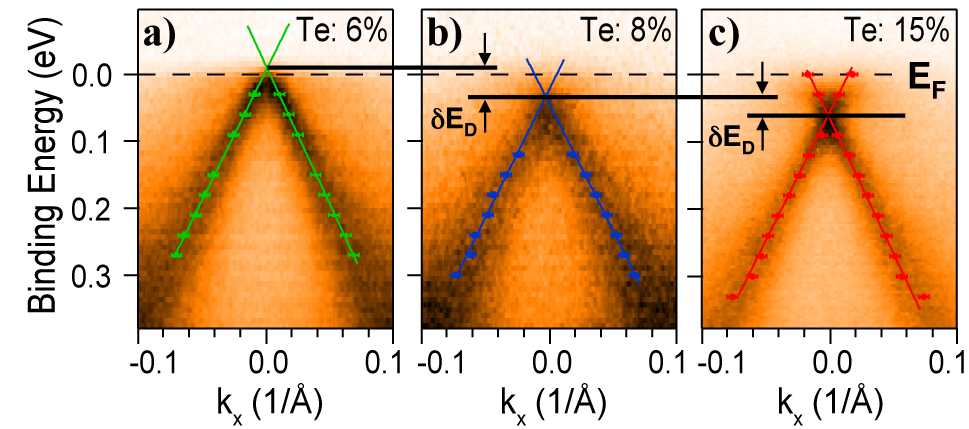}
 	\caption{\label{}Doping control. On varying the Te concentration, we shift the Dirac-point from 10\,meV above \Ef\ (a) for a Te-concentration of 6\% to 30\,meV (b) and 60\,meV (c) below \Ef\ for Te-concentrations of 8\% and 15\%, respectively. Colored lines in a) - c) are linear regressions to the fitted MDC peak positions (colored symbols). All measurements taken along the \GbarXbar-direction, at h$\nu$=21.22\,eV, and at room temperature in our home lab.}
  \end{figure}
Finally, we want to address the role of the Te adatoms during the growth of our samples. As observed by low energy electron diffraction (see supplementary Fig. S2 [25]), the overall surface quality of our samples is increased as Te is added. This indicates that Te acts as a surfactant during the growth of \aSn\ films, similar to findings in Ge epitaxy \cite{Sakata:2000ga}. The benefit of Te for the surface quality is further underpinned by the fact that we where not able to observe the TSS in samples without any Te added as in Fig.\,3 a). We speculate that a minor surface quality smears out the TSS in our photoemission experiment similar to observations in \BiSe\ \cite{Hatch:2011cx}.

Importantly, we find that Te acts as an electron-donor which we demonstrate by dispersions measured on three samples with different Te concentrations (Fig. 5). The concentration is controlled by the Te flux during the growth and determined by core-level intensity ratios of Sn and Te. By increasing the Te concentration from $\sim$6\% (Fig. 5 a) to $\sim$15\% (Fig. 5 c), we can shift the position of the Dirac-point from $\sim$10\,meV above \Ef\ to $\sim$60\,meV below \Ef. We attribute the slight p-doping at low Te concentrations to a diffusion of In atoms from the substrate into the sample.

In conclusion, we have established the 3D topological insulator phase in strained \aSn\ by means of spin- and angle-resolved photoemission accompanied by \textit{ab initio} theoretical calculations. We find good overall agreement to relativistic LDA+U as well as GW quasi-particle calculations. Our results show  that the band order in \aSn\ is different from the known TIs of the \BiX\ family of materials. As a consequence, a surface state emerges out of the second lowest $\Gamma_7^-$-valence band that shows all characteristics of a TSS. 

The close similarity of the \aSn\ bulk band structure to HgTe makes our system a promising candidate to exhibit the quantum spin Hall effect, if the film thickness is reduced to the 2D limit. In this context, it may be important to note that very recently, Sn monolayers were suggested to exhibit the quantum spin Hall insulator phase with a bulk band gap similar to 300\,meV which would allow room temperature applications \cite{Xu:2013ui}. Moreover, the underlying diamond lattice of \aSn\ and the fact that it is the first realization of an elemental topological insulator opens up easier pathways of engineering future devices. 

We are grateful for discussion with Oliver Rader and Laurens Molenkamp. This work was supported by the Deutsche Forschungsgemeinschaft under grants FOR 1162 and SCHA 1510/5-1.

\newpage
\thispagestyle{empty}
\section{Supplementary Information}

\subsection{Theory}
Density functional theory
calculations have been performed employing the full-potential linearized
augmented planewave method as implemented in the FLEUR-code
[1]. DFT calculations in the local density approximation (LDA)
or generalized gradient approximation (GGA) predict the wrong band
ordering (like in HgTe \cite{Sakuma2011}) and a metallic behavior at the
L-point \cite{Rohlfing1998}. To correct these deficiencies, we applied
the DFT+U scheme in the spirit of Cubiotti et al. \cite{Cubiotti1999}
with U = 3.5\,eV. Surfaces have been simulated via calculation of films of
more than 50 atomic layer thickness, to achieve a decoupling of the two
surfaces of the film. Different surface terminations (e.g. a
p(2 $\times$ 1) reconstruction or a p(1 $\times$ 1), hydrogen and tellurium
termination) lead to different (trivial) surface states, but the
topological surface state (TSS) shows up as a resonance in the bulk-projected
states in all cases. 

Different from DFT bandstructures of finite film
setups, the spectral function calculated in a semi-infinite surface
geometry allows a very direct comparison with experimental data. The
infinite bulk attached to one of the surface regions provides a
continuous spectrum in which surface resonances and surface states can
be identified as peaks protruding over the bulk background. We performed
such semi-infinite calculations using the Green function embedding
method \cite{Wortmann2002},  with which a
surface region embedded between a semi-infinite bulk substrate and a
semi-infinite vacuum can be obtained, whose imaginary part yields the
Bloch spectral function of the surface $A(K,e)$. Neglecting matrix
element effects and more complex interactions, the Bloch spectral
function integrated over the surface region is the quantity measured in
ARPES. While the projected bulk states and surface resonances are easily
visualized by plotting the Bloch spectral function, truly localized
surface states can only be plotted after performing broadening by adding
a small imaginary part to the energy due to the infinite lifetime of
such states in a DFT calculation. The result of our semi-infinte calculation is presented in Fig. S1. Two linearly dispersing states, on the bulk band background, are clearly discernible.
 
 \begin{figure}[h]
 	\includegraphics{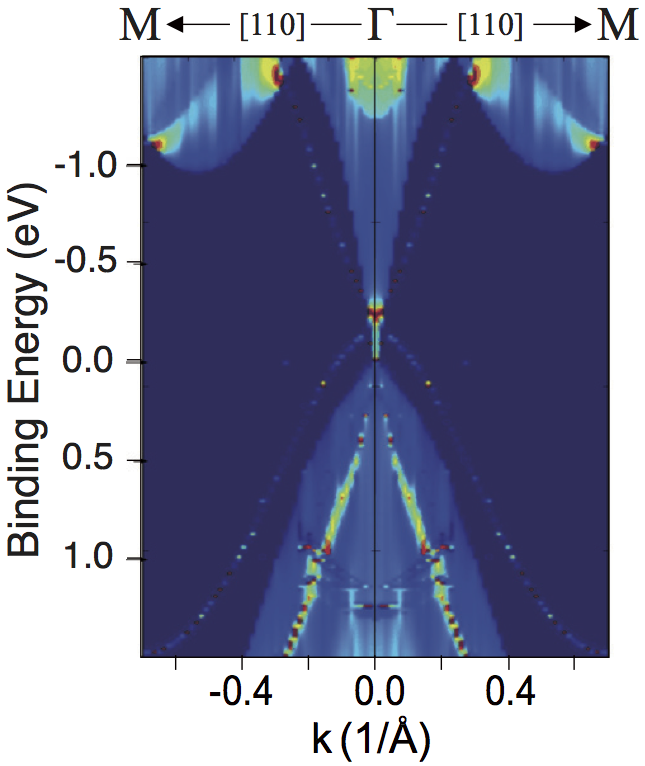}
 	\caption{\label{} LDA+U calculation of 6 layers \aSn\ embedded in a semi-infinite bulk. The TSS is visible as linear dispersion on a weak bulk projection background.}
  \end{figure}

We have also performed many-body calculations within the GW
approximation of the bulk band structure of the cubic and strained \aSn. The GW corrections
to band energies have been added perturbatively to the LDA band structures obtained
within the plane-wave, pseudopotential method, using the Sn$^{4+}$ pseudopotentials.
Special care has been paid to the contribution of the atomic core charge of Sn within 
the GW approximation, as discussed in \cite{2005PhRvB..71d5207F}.
The GW bulk band structures have been parametrized according to the non-orthogonal sp 
tight-binding model, similary as was done in \cite{Pedersen:2010cp}, and the obtained parameters were 
used in calculations of the surface band structure of large slabs containing several 
hundreds of Sn layers. 

\subsection{Experiment}
The growth properties of $\alpha$-Sn on
InSb(001) have been studied extensively in the literature. A seminal
paper was published by Farrow \textit{et al.} \cite{Farrow1981} who reported a
high structural quality from x-ray diffraction up to $0.5\,\mu m$ film
thickness. The $\alpha$-Sn phase in the diamond lattice is the
thermodynamically stable form for bulk material below $13^\circ$C, above
which the solid will transform to the metallic $\beta$-phase without TI
properties. However, the heteroepitaxial film growth serves to stabilize
the $\alpha$-Sn phase up to temperatures of $230^\circ$C
\cite{Mason1992}, so that it is stable for all practically relevant
temperatures, and in addition can be subjected to moderate annealing
treatments if desired. The effect of the growth conditions (rate,
substrate temperature, annealing) to achieve an optimized film quality
is well documented \cite{Farrow1981,Mason1992,Magnano2002}.

We have performed epitaxial deposition studies by using InSb(001)
substrates using commercial wafers. Prior to epitaxy, the surface is
prepared by several cycles of Ar ion sputtering and thermal annealing.
The formation of a surface reconstruction on InSb, i.e., a c(8 $\times $2) superstructure visible in electron
diffraction, is taken as indication for  long-range-order and small defect density. Epitaxy of Sn on a substrate kept at
room temperature yielded the best results, leading to a
(2 $\times$ 1)-reconstructed surface, as reported in the literature
\cite{Farrow1981,Mason1992}, even without
anneal. Incorporating Te as a dopant which also acts as a surfactant resulted in a (1 $\times$ 1)
diffraction pattern of particularly good quality (Fig. S2).
 \begin{figure}[t]
 	\includegraphics{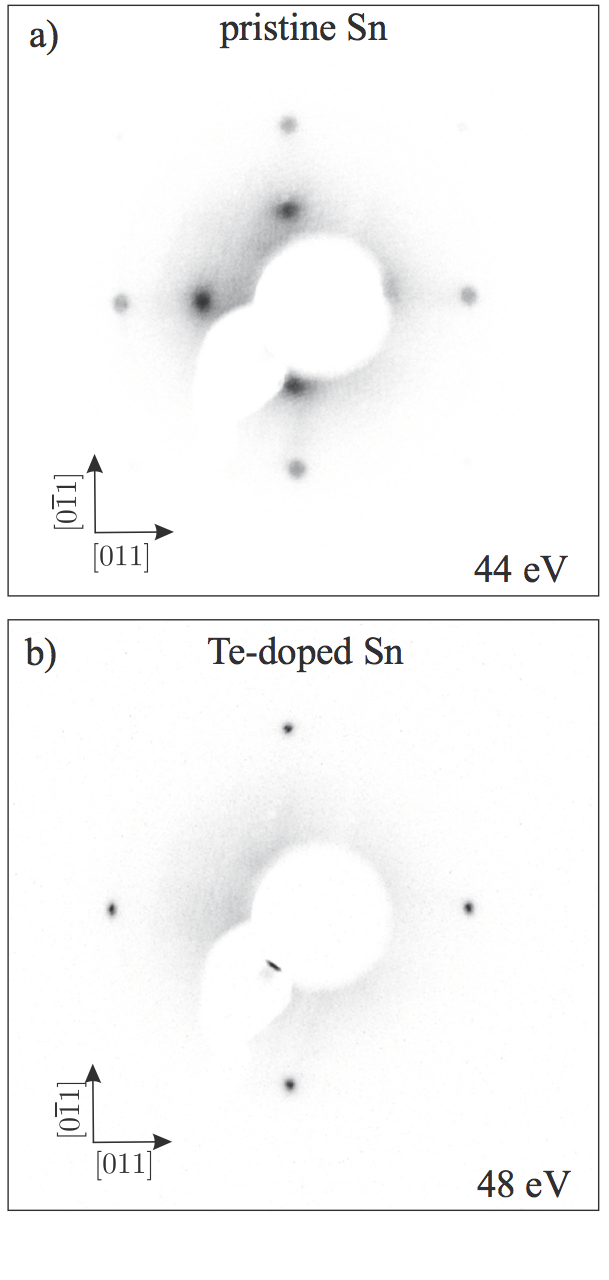}
 	\caption{\label{} LEED patterns of a) an undoped and b) a Te-doped \aSn\ sample. Incorporating Te during the growth leads to an increased surface quality, as obvious from the sharper LEED spots in b) and to a (1 $\times$ 1) surface reconstruction, while the pristine sample in a) is (2 $\times$ 1) reconstructed.}
  \end{figure}
  
Subsequent ARPES measurements  were performed on a six-axis goniometer using
either He-I (21.22\,eV) or Ne-I (16.86\,eV) discharge radiation. The
photoemission data were acquired at room temperature using a SPECS
Phoibos 100 imaging analyzer, typically set to $~30$\,meV and
$~0.2^\circ$ resolution.

Further ARPES measurements were done at beamline 7.0.1 at the Advanced Light Source and at the SIS - X09LA beamline of the Swiss Light Source.
Both endstations are equipped with Scienta R4000 electron energy analyzers.
Spin-resolved measurements have been carried out at the COPHEE endstation at the SIS - X09LA beamline of the Swiss Light Source.

\end{document}